# Beyond 5G: Big Data Processing for Better Spectrum Utilization

Adrian Kliks, Łukasz Kułacz, Paweł Kryszkiewicz, and Hanna Bogucka, Poznan University of Technology

Marcin Dryjański, Magnus Isaksson, Georgios P. Koudouridis, and Per Tengkvist, Huawei Sweden

*This article emphasizes the great potential of big data processing for advanced user- and situation-oriented, so context-aware resource utilization in future wireless networks. In particular, we consider the application of dedicated, detailed and rich-in-content maps and records called Radio Service Maps, (RSM) for unlocking the spectrum opportunities in 6G networks. Due to the characteristics of 5G, in the future, there will be a need for high convergence of various types of wireless networks, such as cellular and the Internet-of-Things (IoT) networks, which are steadily growing and consequently considered as the studied use case in this work. We show that the 6G network significantly benefits from effective Dynamic Spectrum management (DSM) based on RSM which provides rich and accurate knowledge of the radio context; a knowledge that is stored and processed within database-oriented subsystems designed to support wireless networks for improving spectral efficiency. In this article, we discuss context-aware RSM subsystem architecture and operation for DSM in convergent 6G radio and IoT networks. By providing various use-cases, we demonstrate that the accurate definition and access to the rich context information leads to a significant improvement of the system performance. In consequence, we also claim that efficient big-data processing algorithms will be necessary in future applications.*

## Introduction

The demands in capacity and latency for the fifth generation (5G) mobile networks can be summarized as 1000x more capacity of the fourth generation (4G) technologies and latency over the radio link occasionally lower than 1 ms. According to ITU IMT-2020 requirements [1], peak bitrate speeds of up to 20Gbps are expected for 5G. At the same time, in 4G, at frequencies in the Sub-6 GHz domain, the average download bitrates are below 20 Mbps. Verizon, in their millimetre wave 5G trial-fields at 28 GHz in U.S, have achieved throughput speeds of 1.8 Gbps but with a rather low latency of 1.5 ms [2]. 5G systems open high frequencies (up to 30 GHz) to practical application, but these mm-wave frequencies are short in travel distance, requiring lots of radio-heads and antennas to be deployed. Knowing that, one may observe that in 6G the possible operating frequencies may be shifted up to 100GHz or even to 1THz to utilize wider potentially available bandwidths. Moreover, more stringent requirements on latency (i.e., even below 1 ms, down to microsecond level) and rates (i.e., even up to 100 Gbps or 1 Tbps in peak) may be defined. Such Key Performance Indicators have been discussed in [3]. Although base stations working in the terahertz band will operate with less power while providing more capacity (enabling thousands of simultaneous wireless connections), they will require an extreme densification of the network. As a result, massive multiple-input multiple output (M-MIMO) with spatial multiplexing could be subject to further research, so that the bandwidth is spatially reused simultaneously.

In this context, we suppose that for the next, sixth generation (6G) networks even more extreme requirements will be defined than it is currently the case for the 5G network. These requirements will cover various network domains and features, such as higher rate, lower latency (towards, e.g., real tactile internet), high reliability, security, programmability etc. In order to meet the above prospective demands, mobile network operators (MNOs) have to either utilize the available spectrum more efficiently or assign new frequency bands for communication purposes. The latter approach is associated with high cost for attaining long-term transmission licenses. Recent studies have shown that improvements in spectral efficiency can be attained by further densification of the radio access network (RAN), application of massive MIMO and millimeter-wave solutions, and by more flexible spectrum sharing approaches to dynamic spectrum management (DSM) [4], [5]. Various sharing schemes have been investigated so far, just to mention light licencing, pluralistic licensing, co-primary sharing or micro-leasing, among others [6]. Two of them are of particular importance nowadays, i.e., the licensed shared access (LSA) and citizen broadband radio service (CBRS) [6]. We foresee, however, that for 6G networks even more advanced, yet mature solutions will be defined, which will benefit from the access to rich context information, such as [7]-[8].

Thus, efficient and adaptive spectrum sharing and usage becomes a matter of dynamic network planning, and interference coordination among operators' access nodes. This poses demands on the availability of a huge amount of information (also called big data) that determine the intra- and inter-RAN dependencies. Today's technologies are mainly focused on radio service-related maps (RSMs). These maps are populated with data related to radio aspects as in the well-known radio environment maps (REMs), such as coverage maps or interference maps. On top of that, they contain various types of network- and service-related data, such as traffic maps (distribution of requested or served traffic, traffic patterns), or trajectory databases (mobility patterns of users) [9]. In [10] the authors considered the usage of REM, a kind of RSM, for reducing the number of handovers by predicting the trajectory of the users. Knowing the channel characteristics (stored in REM) and the prospective route of the user, the system procedure of the hand-off may not be initiated if the so-called ping-pong effect is expected. However, the more data is required by the DSM operation, the higher the operational costs and the load on the backhaul part of a system. For this reason, current technologies restrict the RSM to statistical databases which are proven suitable mainly for long-term system optimization [9]-[10]. Particularly, in [11] the authors discussed the ways how in general the artificial-intelligence-oriented algorithms could improve 5G radio access networks.

Future 6G wireless networks may significantly benefit from the access to rich and accurate context information, but this entails the need for highly efficient processing and management of these big-data sets. Our considerations in this paper are also motivated by the fact that various types of context information are typically available in wireless systems, and therefore, the performance of numerous algorithms for signal processing and communication can be improved. This is because many algorithms utilize feedback information about the communication channel or user preferences to better adjust to the ongoing scenario.

The contributions of this article are the following:

- First, we discuss our generic observation on the prospective processing of massive-information in future wireless networks
- Next, we emphasize how this idea can be tailored to radio-oriented schemes and propose a solution of the RSM-based subsystem as part of the future 6G architecture
- We evaluate the benefits of the proposed solution in various scenarios showing how the 6G network may outstrip 5G.

In Section II, we discuss a context-aware RSM subsystem and an architecture for DSM in convergent 6G radio and IoT. In Section III, we discuss an RSM subsystem implementation example, and demonstrate that the accurate definition and access to the communication context can lead to a significant improvement of the system performance. We evaluate the context-aware RSM-based approach based on an inter-RAN interference scenario where multiple 5G operators cooperate to maximize the system throughput. Finally, in Section IV, we address the issue if the RSM architectural and algorithmic framework can unlock the spectrum potential for future IoT and 6G radio, and we provide conclusions in Section V.

## Dynamic Spectrum Management for Convergent 6G and IoT Networks

### Beyond 5G

Advanced database technologies constitute the foundations of numerous contemporary technological solutions, especially in the area of computer science and communications. However, we would like to differentiate between the current approaches (applied in contemporary networks or to be implemented in NR systems), and the proposed RSM-based subsystem which is discussed below.

Databases of various types (relative, graph, etc.) are widely used nowadays in many contexts. On the one hand, one may consider the registers for storing user-specific data (such as a Unified Data Management entity in the NR architecture), on the other – advanced technologies designed for processing and computing at the edge, in the fog or cloud. What is, however, envisaged for 6G, is the presence of a dedicated map-oriented subsystem created and managed for collecting, storing and processing of massive-information about the communication context. In other words, it includes all the information about the context (such as geographical data, architectural maps, telecommunication data, electromagnetic data, but also user behaviour in different time scales, habits, user profile, social data, data from other systems, such as public transportation, weather forecasts, etc.) that might be useful for improving the performance of the network. In order to populate the database and to keep it up-to-date, the information has to be gathered following the crowd-sourcing approach. Databases may be shared between operators or may be in the sole possession of one stakeholder, however, in the former case, standardized interfaces have to be defined. The volume and variety of kinds of prospective information entail the need for customized processing typical for big-data schemes.

One may consider the following analogy: what we foresee for 6G in terms of massive context-information processing with regard to 5G solutions, is similar to the technological and conceptual change between the conventional MIMO scheme in LTE and its massive version in NR. In what follows, we focus our analysis on radio-specific aspects of the proposed solution, i.e., on the application of an RSM-based subsystem for context-aware communications.

*Context-Aware Communication*

Originally, the term "context-aware communication and computing" referred to a software engineering solution which allows for the automatic customization of devices, systems, and applications according to the changing context of the radio environment, e.g., [12]. The context refers to the information characterizing the operational state of each entity in MNO's RAN. It comprises a set of parameters that define the state of RAN, in detail. Typically, location information (a map) is associated with a set of context parameters that define certain aspects of RAN states. Numerous types of maps can be identified at different hierarchy levels, ranging from the level of base stations (BSs) to the level of RAN, backhaul and core network. At each level of the hierarchy, different contexts can be utilized to characterize the corresponding operational states, i.e., higher level description can utilize less specific, generalized metrics/parameters. The hierarchical approach aims at addressing the problem of complexity and computational cost, which in dense heterogonous networks is expected to be considerably high. This is mainly due to an increasing amount of prospective context information requiring big-data processing with high geographical granularity. Moreover, artificial intelligence algorithms are envisaged for proper inference and reasoning based on rich context information. Please note that by referring to big data processing we refer to the definitions and approaches presented in [13] where the datasets are described by four or five "Vs", i.e., volume, velocity, variety, veracity and value.

*Environment and radio service maps*

Numerous types of radio service maps (databases) can be identified. We foresee that the most relevant ones for future networks are the following:
- Radio-data maps of parameters strictly related to the radio environment; these can be divided into: *i)* maps of capacity-related parameters, such as the traffic load, spectrum efficiency, outage probability, etc.; *ii)* maps of coverage-related parameters, including signal-to-interference-and-noise ratio maps, received signal strength indication distribution, etc.;
- Performance maps of system performance indicators, e.g., QoS/QoE, end-to-end transfer time, latency, packet jitter distribution, etc.;
- Mobility maps, containing the knowledge on handoff regions, handover regions, pilot coverage areas, etc.,
- Traffic-density maps, identifying regions of high and low traffic densities;
- User-density maps, containing the knowledge on estimated users density (beneficial for scheduling and admission control algorithms, better load balancing and traffic offloading);
- Trajectory and velocity maps of mobile cells or users moving along typical routes;
- History maps, e.g., characteristics of long-term user behavior or the effectiveness of past spectrum-assignment decisions.

Note that the idea of databases applied for the improvement of system performance has been considered in numerous contexts, in particular, with reference to dynamic spectrum access and spectrum sharing, where radio-service data are of particular importance. In essence, the radio service map (RSM) can contain all of the above-listed maps for radio-service performance optimization in a given area. The RSM-based subsystem for dynamic spectrum management consists of three main functional blocks:
- Databases (repositories), which contain various types of data: private or public, understandable by a human and by a machine,
- RSM manager, playing the role of a cognitive engine, responsible for database management and data processing, overall reasoning and decision making, conformance verification, coordination and control, as well as communication with other RSM managers and corresponding operators' legacy entities,
- Data acquisition function responsible for collecting data from the environment covered by the RSM, either periodically or in an event-driven way. (Here, the application of sensing/monitoring modules can be considered.)

*Architecture of RSM-based subsystem*

The architecture of a generic RSM-based subsystem containing the above elements is shown in Figure 1. By assumption, each operator has its own RSM-based spectrum management system. This is due to the fact that each operator will be in possession of such user data that cannot be shared, but can be processed for service delivery (e.g., some sensitive user data). The set of repositories is split into two parts: private (including sensitive information) and shared one. Moreover, each RSM-based spectrum management system needs dedicated interfaces between the RSM and Network domain. The operators' RSM subsystems may exchange data directly, however, the publicly available data can also be managed by a 3rd-party RSM subsystem. The generic RSM-subsystem structure may be further adjusted for specific applications.

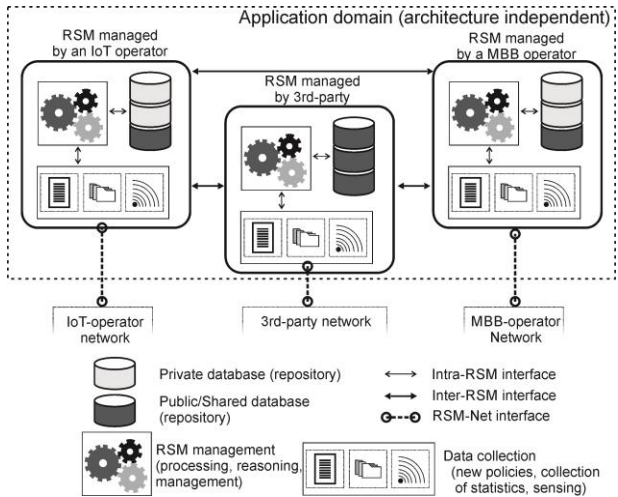

***Figure 1.*** *The structure of the database system at the application domain.*

The RSM subsystem can be implemented either in a distributed or centralized manner, and this implementation depends on the deployment of BSs in the covered area. If the BS density is low, it should be beneficial to use a centralized RSM for their coordination. The physical location of the RSM subsystem elements depends on the application time scale for which the database is designed. Thus, a layered (hierarchical) architecture seems to be most appropriate for practical deployment, where long-term data are stored in centralized RSMs, whereas short-term information is saved in distributed, local RSMs.

### Defining the Context

Rich-in-content and possibly massive datasets defining the surrounding environment (i.e., widely understood communication context) can improve the performance of the future network. This will be possible when the quality and veracity of the data is preserved, as well as the access and data processing is manageable. In Table 1 we compare selected aspects of context maintenance.

*Table 1. Summary of the challenges for developing clear context dataset*

| Parameter | Discussion referring to the DSM |
|---|---|
| Data collection and their veracity | Challenge: Massive datasets have to be updated, this requires permanent (periodic or continuous) data gathering |
| | Supporting technology/solution: Crowdsourcing achieved by the deployment of dedicated sensors and/or utilization of measurements conducted by UEs |
| Variety of data | Challenge: In terms of dynamic spectrum access, numerous maps can be defined; depending on the application use case, some maps will be of higher importance |
| | Supporting technology/solution: Although the types of created and stored maps have to be decided in the design phase, the role of each map may vary in time, so adaptive algorithms could be applicable |
| Storage and processing | Challenge: Due to the "five Vs" concept [13], massive datasets require specific approaches for their storage and processing. |
| | Supporting technology/solution: Currently developed technologies like fog/edge/cloud computing are the basis for the development of new frameworks for big data processing for DSM |
| Ownership | Challenge: Management of a dedicated RSM-based subsystem is costly |
| | Supporting technology/solution: Various databases may be shared between the operators reducing OPEX; also, the establishment of dedicated context-information datacentres (as it is now the case for cloud-computing) could be a valuable option |

## Performance of Context-Aware Dynamic Spectrum Management

### Where 6G radio and IoT overlap – example scenario considerations

Let us consider an example scenario with two operators (see Figure 2). Mobile IoT-network operator deployment aims at delivering machine-type communication (MTC) [1] traffic services to users located outdoors. It is provided by dedicated road-side units operating at frequency $f_1$, and their coverage areas are depicted in green in Figure 2 b). The coexisting mobile broadband MBB operator delivers MBB services to both indoor and outdoor users. The outdoor users are primarily served by macro-BSs working at frequency $f_2$ (depicted in red). However, as a result of frequency $f_1$ not being utilized indoor, the set of indoor hotspots is deployed in the band originally assigned to the IoT operator to improve the end-user performance (coverage depicted in orange). The MBB operator deploys multisystem 4G/5G/6G BSs, denoted for simplicity and backward compatibility as eNB/NR (new radio) BSs. Although the investigation has been carried out for the case where indoor operator may deploy multi-system nodes (allowing for switching between the systems), the results presented in this work concentrate intentionally on the case when indoor operator deploys NR base stations.

---

[1] We use the terms MBB and MTC to preserve backward compatibility with 4G and 5G networks' nomenclature, however in the context of 6G network new schemes may be defined

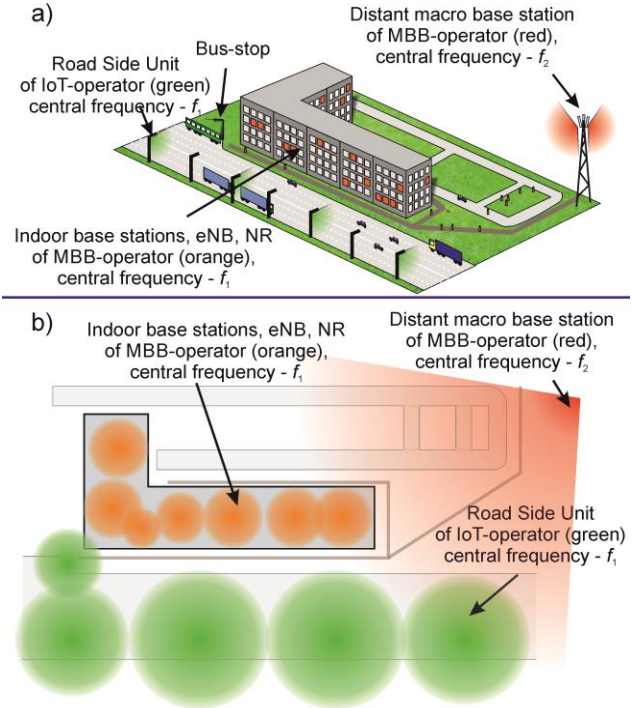

*Figure 2. Example scenario: a) 3D deployment view, b) Top view of the radio coverage*

## Use-cases and setup

In order to evaluate the performance of the proposed context-aware communications with the spectrum shared between IoT and MBB networks, we have considered three use cases:

A. Real-time LSA spectrum sharing: Application of LSA policies for the considered scenario, where the policies define the spectrum sharing rules between both operators
B. Outdoor user interference reduction: Application of the RSM-based subsystem where there is an outdoor user moving along the building (in this case the trajectory, history and handover maps are applicable)
C. Advanced spectrum management: Application of traffic maps for better spectrum management by the MBB operator through advanced scheduling under constraints introduced by the IoT operator.

Below, we provide some representative system-level simulation results for these use-cases. As the technical details for 6G are not available at this stage, for simulation purposes, we consider a setup tailored to 4G and 5G to illustrate the expected gains and benefits. In general, the simulations represent at least 1 second, with scheduler decisions every 1 ms. The inter-cell interference cancellation algorithm has been applied (a version of soft frequency reuse scheme), and Winner II channel models have been used. Arbitrarily, the central frequencies have been set to $f_1$ = 3.5 GHz (typical for 5G systems, also called NR – New Radio), and $f_2$ = 2.6 GHz (for conventional systems, i.e., LTE-A). The multicarrier SISO communication of 20 MHz channels has been simulated. Both cooperating systems (IoT and MBB operators) follow the LTE-A/NR specific spectrum access schemes, i.e., the standardized modulation and coding schemes are selected based on effective exponential SNR mapping (EESM), and implemented jointly with the proportional fair (PF) scheduler (for which the *β* parameters used for weighting instantaneous and averaged rate of each user was set to 0.5). Please note that at the time of conducting the investigation not all of the physical-layer specific parameters of the NR have been known, thus, some assumptions have been done, as stated in Table 2. All the simulations have been carried out in the Matlab environment.

*Table 2. Simulation parameters used in experiments*

| Parameter | Value / notes |
|---|---|
| Path loss channel model | Winner II channel model for indoor/ indoor-outdoor propagation schemes; |
| Channel changes | Extended Pedestrian A model (EPA) with 7 tabs implemented by means of Jakes model |
| Central frequency | $f_1$ = 3.5 GHz (NR) and $f_2$ = 2.6 GHz (LTE-A) |
| Channel bandwidth and its utilization | 20 MHz / 98% (as for NR) |
| Corresponding number of RBs | 108 (as for NR) |
| Scheduling method | Proportional Fair with $\beta$=0.5; decisions made every 1 ms |
| SNR mapping | EESM tailored appropriately to NR with 15 MCS levels |
| Maximum transmit power | 21 dBm |
| BS height | 10 m (above ground for outdoor micro BS), below ceiling (for indoor BS) |
| Minimum distance between UE and BS | 10 m (for outdoor BS), 3 m (for indoor BS) |
| User distribution | Uniform, 20% of static users (no movement), 80% of pedestrian users (approx., speed 0.8 m/s) |
| User Data Traffic | Full Buffer mode |
| NR – New Radio (5G) EESM –Exponential Effective SNR Mapping LSA – License Shared Access MCS – Modulation and Coding Scheme RB – Resource Block | |

## Application of the LSA changing in time

In this considered use-case, the database can be utilized to adapt to the changes in the LSA policy in real-time. Assume that the MBB operator borrows a certain fragment of spectrum from the IoT operator for a given period of time. Once this period is over, new policies are applied. (The RSM subsystem repository includes the policy schedule.) Each record in the database defines the ID of an indoor BS, the time-range when the frequency band can be used by the BS, the maximum transmit power and other transmit power constraints, and the frequency-range that can be used. For demonstration purposes, we have defined an arbitrary sequence of frequency changes.

Figure 3 shows the achieved rate variations in time as functions of changes in the assigned bandwidth for indoor BSs. At the bottom of the figure, the average number of resource blocks available for indoor-BSs of MBB operator is plotted as reference, to justify the time-changes in indoor-user rate. The results show that the application of the RSMs allows for real-time configuration of BSs. Thus, one may conclude that it is possible to apply very frequent changes in terms of frequencies assigned to BSs; thus, adaptive frequency allocation provides new prospective benefits for performance.

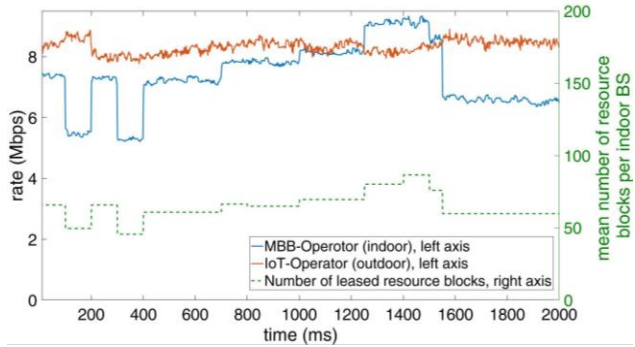

*Figure 3. Average rates for indoor and outdoor UEs as the function of time in the advanced LSA.*

## Application of the RSM databases with an outdoor user moving along the building

Let us now evaluate the applicability of RSMs for controlling interference generated to the moving outdoor user associated with the IoT operator (it may be a pedestrian or a bus moving along the street as in Figure 2). One may observe that when the bus/pedestrian is close to the building corners, indoor BSs deployed close to these corners have the strongest impact on the induced interference. The influence of BSs deployed in the central part of the building are smaller, and there is practically no influence of BSs located in the opposite corner of the building. At the same time, when the moving user is close to the central part of the building, it observes interference coming from the highest number of indoor BSs.

In this use-case, the optimization algorithm [14] takes the trajectory map of the outdoor user into account, and finds the transmit power of all indoor BSs to maximize the sum-rate, while keeping the interference induced to the outdoor user below the agreed level. This result is compared with the situation when there is no indoor network. In our demonstration scenario, we consider a bus driving at 50 km/h along the building, which takes 6 seconds. The mean indoor/outdoor UE rate is calculated over periods of 200 ms. The achieved mean rates are shown in Figure 4. While analysing the results, please notice that we have intentionally limited the number of outside UEs to one – this allows for the evaluation of the impact of this particular UE on the overall system performance.

- There is a slight degradation in the achievable rate for outdoor users due to spectrum sharing (see the results denoted in the legend as "Adaptive outdoor protection");
- When there is no protection of the outdoor users, there is a strong degradation of the outdoor-user rate as a function of location, whereas the indoor-user rate is unchanged;
- Because the optimization algorithm requires the knowledge on the outdoor user position, the mean outdoor rate is unchanged regardless of the outdoor user position in the "Adaptive outdoor protection" scenario. In the same time, the rate of the indoor-users is the worst when the outdoor-user is close to the central part of the building where indoor-BSs power is limited.

## Application of traffic maps for better spectrum management

Let us now consider the use-case in which the RSM subsystem is applied with the traffic maps for the purpose of resource scheduling simplification. In the baseline LTE (4G) system, the scheduler makes its decision every 1 ms [14]. Such frequent decision making is not necessary if the scheduler can predict the behavior of a user. Particularly, it is beneficial in the office scenario, where the users (or IoT devices) do not move in a period much longer than 1 ms. In our use-case scenario, apart from static or slowly moving users, we consider an MBB

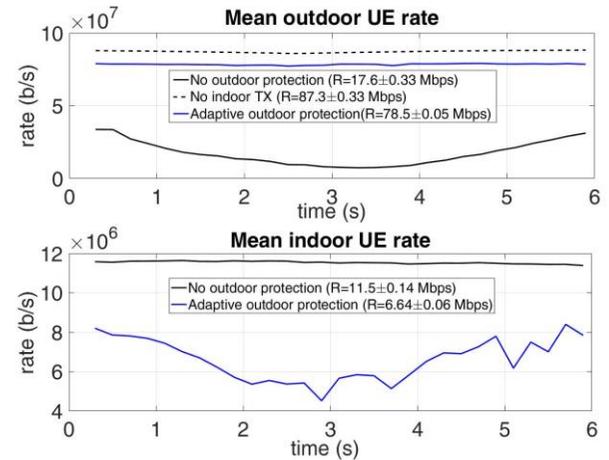

*clustered* user consisting of 10 non-moving devices, resulting

*Figure 4. Mean indoor and outdoor UEs rates vs. time (one outdoor UE).*

in relatively slowly-changing channel coefficients. The following algorithm has been implemented:
- In the initialization step, an observation period is applied for about 10 seconds, during which the RSM subsystem collects the decisions made by the scheduler on each resource block;
- Then, the RSM manager algorithm determines the average modulation-and-coding scheme (MCS) index assigned to each resource block for a certain location, and calculates the standard deviation of the assigned MCS index; the MCS average and the standard deviation maps are shown in Figure 5. The historical decisions and the current ones are weighted in order to smoothly replace older decisions with updated ones;
- Once the map is created, the algorithm selects the set of static devices (with the smallest standard deviation and non-zero mean) and assigns the resources to these UEs, so that their traffic demands are satisfied. In consequence, the set of resources is split into two parts: one assigned to the static users for longer time (5 seconds in our simulations), and the other – assigned to the moving or new users based on the traditional PF algorithm [15]. After 5 seconds, the algorithm checks if there are any changes in the currently observed traffic distribution and adjusts its behaviour to the new situation.

In summary, access to this kind of context information (e.g., the traffic map) allows the scheduler to reduce the number of its scheduling decisions by simply omitting static users in the dynamic resource allocation phase. Static users are allocated with previously defined parameters.

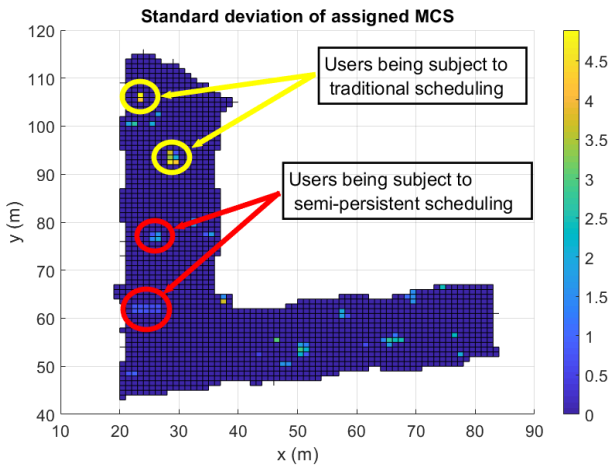

*Figure 5. Traffic map – the standard deviation of the MCS index after 15 s of observation; each point represents 1 m x 1 m square.*

In Figure 6, the sum-rates for indoor and outdoor users are shown. As expected, there is no change in the achieved rate for both indoor and outdoor UEs. The complexity of the scheduler is reduced without any negative impact on the system performance. Observe the dashed vertical line denoted as "Update", indicating the periodic update of the decision on the assignment of users to two scheduler groups. The average number of resources allocated by the PF scheduler to indoor users is reduced from 510 in the initial phase to 446 in the second phase (12.5% reduction of the processing cost). Such a reduction means a huge gain in the entire processing in the MBB system for the same outdoor system performance.

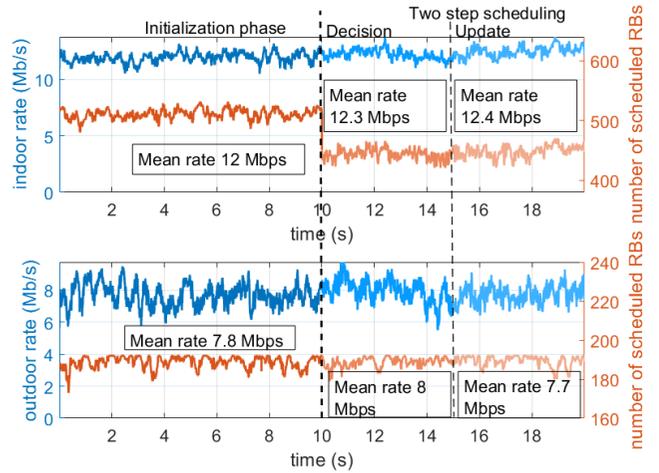

*Figure 6. Sum-rates achieved for indoor and outdoor UEs using RSMs with traffic prediction.*

## Discussion on RSM and Database Subsystem

The presented discussion and results show the benefits of context-information availability in databases. However, the trivial conclusion that *the more information available, the more opportunities for efficient communication*, is not that straightforward in practical implementation. It entails challenges for wireless communications which stem from the domain of big data processing, especially if mobility aspects are considered. Access to the accurate and valid data in dynamically changing environments emphasizes the need for efficient design of the RSM subsystem, where all potential delays in collecting data are minimized. Moreover, the statistical databases should be replaced with more advanced solutions tailored to such rapidly varying scenarios. At the same time, a big amount of data to be available in various locations implies either information caching (to reduce delay) or high backhaul traffic. One may foresee that once the availability of rich context information is guaranteed, numerous algorithms from various domains of wireless communications can be enhanced for better performance. The concise summary of the design aspects of the RSM is sketched in Table 3. Moreover, in Table 4 we have presented brief analysis of the prospective load from the operator perspective due to the application of the RSMs for each of the considered use cases. As the real load is highly application and implementation specific, we have presented the worst-case analysis, i.e., no storage optimization is applied, no advanced data structures were considered.

Table 3. Selected aspects of the RSM subsystem design

| RSM design issue | Notes on implementation |
|---|---|
| RSM as a separate subsystem | Contemporary networks have already implemented procedures for data collection and utilization. The trend towards the application of big-data solutions and the need for data exchange with other systems and operators prompts a separate RSM subsystem to be |

| RSM design issue | Notes on implementation |
|---|---|
| | created, as it is done today for an operations/business support system (OSS/BSS). |
| RSM management | RSM subsystem functionality can be delivered by a company specialized in data collection, environment monitoring, management, and storage, i.e., the network operator can outsource such tasks to companies delivering data processing services. |
| The need for standardization of RSM databases | The ways of collecting of some specific types of data may be subject to standardization (e.g., feedback information about the channel state). However, each operator may propose its own tailored solutions for better data aggregation and storage. Nevertheless, there should be an interconnection between various systems and operators (e.g., for better gathering of data about the ambient environment), and the communication interfaces should be defined. |
| Centralized or distributed architecture | The advantages and disadvantages of centralized and distributed architectures are well investigated. Considering the specific features of a wireless network (e.g., user mobility, multipath propagation, and large scale of the communication networks), the hybrid solution could be the best option, where long-term data are stored centrally, whereas short-term information are cached locally. |
| Consistency and availability | Following the well-known CAP theorem on consistency, availability and partition tolerance, the design of the distributed database should take these aspects into account. |
| Delay and accuracy | In mobile wireless networks, the key functional problem is the reliable access to accurate data exactly at the right time (measured in milliseconds). The requirements for end-to-end delay constrain the RSM subsystem architecture. It may be required to apply advanced data processing and data mining algorithms for efficient data delivery. |
| Data gathering | The RSM subsystem is beneficial only when it is populated with updated data. Thus, methods of efficient monitoring and storage must be implemented. Standalone sensors may complement the information collected within the wireless network by mobile agents or BSs/access points. |
| X-Haul traffic | The presence of an RSM subsystem responsible for big-data processing in wireless networks immediately generates the need for efficient front- and back-haul (X-haul) traffic management. This aspect is directly related to the subsystem architecture design, whether centralized or distributed. However, access to big data also assumes that there is a dedicated communication channel through which the context data are transferred. |
| Security | Collection and storage of big data requires the application of advanced security algorithms for better privacy protection. The anonymization of collected data is a good approach, however, it has to be supported by proper security procedures. |

Table 4. Storage and AI tools analysis (worst case scenario)

| Use case | Brief Load Analysis (Worst Case analysis) |
|---|---|
| *Note:* we map each use case to one or more of the Seven Patterns of AI, which consists of: 1. Recognition, 2. Conversation and Human Interaction, 3. Predictive Analytics and Decisions, 4. Goal-Driven Systems, 5. Autonomous Systems, 6. Patterns and Anomalies, 7. Hyper-Personalization. | |
| A. Real-time LSA spectrum sharing | Storage: In this case, the leasing operator specifies the rules (i.e., allowed transmit power per location and resource block – 8 bits, spectrum masks - 40 bits) for spectrum sharing. For the defined number of base stations (12) and resource blocks (100), and considering that the change of configuration may be done every ms, the required storage speed is around 57 MB/s (i.e., such amount of data has to be updated in RSM every second). This value generally depends on the scenario assumptions, and its optimum is implementation-specific.<br><br>AI algorithm: As the general policy rules for spectrum sharing are prepared by humans in human-understandable way (may be done by system administrator), and rules may be written in fuzzy way, the considered algorithms use declarative programming from one hand side, and fuzzy logic on the second side. From the seven patterns of AI this set falls within Conversation and Human Interaction group. |
| B. Outdoor user interference reduction | Storage: Here the policy rule is fixed, so can be omitted in calculations. Moreover, the trajectory map which defines the typical routes of the bus, is also negligible in the overall load. In this case, in the assumed worst-case scenario, the heaviest burden is the transfer of the values of the observed signal quality for each resource block. Even, assuming only 5 bits per resource block (following CQI reporting procedure), the RSM will need to receive 0.5 Mbps per user, then process around 6 Mbps (in total for all 12 BSs), and send back to each secondary user around 0.5 Mbps. The above values scale proportionally with bandwidth (number of resource blocks) and are limited by implementation-specific constraints<br><br>AI algorithm: In this case the clue is to predict the behaviour of the outdoor-user trajectory, and to utilize this information for better spectrum sharing. After the learning phase (training), the typical values of allowed power and resource allocation can be done, so pattern matching schemes are promising. Within the seven patterns of AI the algorithms here fall into predictive analytics and decisions, as well as to pattern detection. |
| C. Advanced spectrum management | Storage: In this case, the RSM utilizes the map of the building (separate per each floor) with a given raster. It has to store and process the historical MCS value per each location, as well as its variance. Assuming raster of 1 dm, 5 bits per MCS index, and 8 bits per variance, it will need only around 21 kb per each resource block, so around 2.1 Mb per all resource blocks. These values will increase accordingly with bandwidth. The processing speed of data stored is at analogous level (i.e., around 2 Gbps).<br><br>AI algorithm: As the locations of the users may change in time, the AI algorithms applied falls again in the pattern and anomaly recognition group within the seven patters of AI. |

# CONCLUSIONS

As in the context of 5G networks, the sixth-generation networks are envisioned to support a large number of services which pose divergent requirements. Collectively, these services are expected to demand more and more capacity, while the networks are required to be more energy-efficient. As an enabler for the converged multi-operator and multi-service networks, big-data processing for context-aware communication with radio service maps is considered to play the key part. In this article, we have presented an application of RSM for dynamic spectrum management in future, context-aware communication networks. Taking the radio service context and operators' priorities into account, we have shown that, complementary to long-term mechanisms that benefit from legacy applications, RSM also shows high value for mid- and short-term RRM schemes. The combination of radio-environment-, traffic- and trajectory maps with prediction functionalities offers a huge potential for efficient resource management and flexible sharing of resources between operators according to their priorities, needs and dynamic changes of the spectrum usage scenarios. Thus, we claim that 6G systems may strongly benefit from the application of big-data processing algorithms realized by means of dedicated, context-information database subsystems.

## Biographies:

*Adrian Kliks (adrian.kliks@put.poznan.pl) is an assistant professor at Poznan University of Technology's Department of Wireless Communications, Poland. His research interests include new waveforms for wireless systems applying either non-orthogonal or noncontiguous multicarrier schemes, cognitive radio, advanced spectrum management, deployment and resource management in small cells, and network virtualization.*

*Łukasz Kułacz (lukasz.kulacz@put.poznan.pl) is a Ph.D. student at the Department of Wireless Communications at Poznan University of Technology, Poland. His research interests include wireless communication physical layer design.*

*Paweł Kryszkiewicz (pawel.kryszkiewicz@put. poznan.pl) is an assistant professor at Poznan University of Technology's Department of Wireless Communications, Poland. His research interests include the cognitive-radio physical layer, multicarrier signal designs for green communication, and interference limitation in 5G systems.*

*Hanna Bogucka (hanna.bogucka@put.poznan.pl) is a full professor and the Wireless Communication Department Head at Poznan University of Technology. Her research area is radio communication: radio resource management, opportunistic radio access, adaptive and cognitive radio.*

*Marcin Dryjanski holds Senior IEEE Membership since 2017. He was a Workpackage Leader in FP7-5GNOW project. Marcin served as a RAN specialist at Huawei working on architecture towards 5G. He specializes in PHY/MAC/ RRM, is a co-author of research papers on 5G design, and a book: "From LTE to LTE-AdvancedPro and 5G".*

*Magnus Isaksson (magnus.isaksson@huawei. com) received his M.Sc and Ph.D degrees in Telecommunication Theory from KTH, Sweden in 1984 and 1990, respectively. He joined Ericsson in 1984 working with mobile network development,*

*radio research, and product management. Since 2013 he is with Huawei as Lab Director.*

**Georgios P. Koudouridis** *(george.koudouridis@ huawei.com) is a principal researcher in the Wireless Network Algorithms Lab of Huawei Technologies Sweden. He received his Ph.D. degree in Telecommunications from KTH, Sweden, 2016. Prior to joining Huawei in 2008, he was with Telia Research. His research interests include radio resource management, and self-organized networks.*

**Per Tengkvist** *(per.tengkvist@huawei.com) is a senior System Architect at Huawei, with more than 20 years' experience of R&D for GSM, UMTS, LTE and since several years 5G. The main research interest is Massive MIMO. Prior to working with Huawei, he was with Ericsson and Sectra.*